\documentclass[structabstract]{aa}  
\pdfoutput=1
\usepackage{graphicx}
\usepackage{txfonts}
\usepackage{natbib}
\bibpunct{(}{)}{;}{a}{}{,} 
\usepackage{color}
\usepackage[labelfont=small,bf,font=small,labelsep=period]{caption}
\usepackage{balance}
\newcommand{\mch}{\ensuremath{\mathrm{M}_\mathrm{Ch}}}
\newcommand{\msun}{\ensuremath{\mathrm{M}_\odot}}
\newcommand{\nuc}[2]{\ensuremath{^{#1}\mathrm{#2}}}
\newcommand{\kms}{\ensuremath{\mathrm{km}\,\mathrm{s}^{-1}}}
\usepackage{hyperref}
\hypersetup{colorlinks,bookmarksopen,bookmarksnumbered,citecolor=blue,
	urlcolor=black,linkcolor=blue,pdfstartview=FitH}
\def\mhref#1{\href{mailto:#1}{#1}}
\usepackage[figure,figure*]{hypcap}

\begin{document}
\title{Gamma-ray diagnostics of Type Ia supernovae}

\subtitle{Predictions of observables from three-dimensional modeling}

\author{A.~Summa\inst{1} \and A.~Ulyanov\inst{2} \and M.~Kromer\inst{3}
  \and S.~Boyer\inst{1} \and F.~K.~R\"opke\inst{1} \and
  S.~A.~Sim\inst{4} \and I.~R. Seitenzahl\inst{1,3 } \and
  M.~Fink\inst{1} \and K.~Mannheim\inst{1} \and R.~Pakmor\inst{5} \and
  F.~Ciaraldi-Schoolmann\inst{3} \and R.~Diehl\inst{6} \and
  K.~Maeda\inst{7} \and W.~Hillebrandt\inst{3}}

\institute{Institut f\"ur Theoretische Physik und Astrophysik,
  Universit\"at W\"urzburg, Campus Hubland Nord,
  Emil-Fischer-Str. 31,\\
  D-97074 W\"urzburg, Germany\\
  \email{\mhref{asumma@astro.uni-wuerzburg.de}} \and School of Physics,
  University College Dublin, Belfield, Dublin 4, Ireland \and
  Max-Planck-Institut f\"ur Astrophysik, Karl-Schwarzschild-Str. 1,
  D-85741 Garching, Germany \and Research School of Astronomy and
  Astrophysics, The Australian National University, Mount Stromlo
  Observatory, Cotter Road, Weston Creek, ACT 2611, Australia \and
  Heidelberger Institut f\"{u}r Theoretische Studien,
  Schloss-Wolfsbrunnenweg 35, D-69118 Heidelberg, Germany \and
  Max-Planck-Institut f\"ur extraterrestrische Physik,
  Giessenbachstra{\ss}e, D-85748 Garching, Germany \and Kavli
  Institute for the Physics and Mathematics of the Universe
  (Kavli-IPMU), Todai Institutes for Advanced Study (TODIAS),
  University of Tokyo, 5-1-5 Kashiwanoha, Kashiwa, Chiba 277-8583,
  Japan }

\date{Received xxxx xx, xxxx / accepted xxxx xx, xxxx}

 
\abstract
{Although the question of progenitor systems and detailed explosion
  mechanisms still remains a matter of discussion, it is commonly
  believed that Type Ia supernovae (SNe Ia) are production sites of
  large amounts of radioactive nuclei. Besides the fact that the
  gamma-ray emission due to radioactive decays is responsible for
  powering the light curves of SNe Ia, gamma rays themselves are of
  particular interest as a diagnostic tool because they provide a
  direct way to obtain deeper insights into the nucleosynthesis and the
  kinematics of these explosion events.}
{We study the evolution of gamma-ray line and continuum emission of SNe Ia
  with the objective to analyze the relevance of observations in this 
  energy range. We seek to investigate the chances for success of future
  MeV missions regarding their capabilities of constraining intrinsic 
  properties and physical processes of SNe Ia.}
{Focusing on two of the most broadly discussed SN Ia progenitor
  scenarios -- a delayed detonation in a Chandrasekhar-mass white
  dwarf (WD) and a violent merger of two WDs -- we use
  three-dimensional explosion models and
  perform radiative transfer simulations to
  obtain synthetic gamma-ray spectra. Both chosen models produce the
  same mass of $^{56}\mathrm{Ni}$ and have similar optical properties that are in
  reasonable agreement with the recently observed supernova SN 2011fe. We examine
  the gamma-ray spectra 
  with respect to their distinct features and draw connections to certain 
  characteristics of the explosion models. 
  Applying diagnostics, such as line and hardness ratios,
  the detection prospects for future gamma-ray missions with higher
  sensitivities in the MeV energy range are discussed.}
{
  In contrast to the optical regime, the gamma-ray emission of our two chosen models
  proves to be rather different. The almost direct connection of the emission of
  gamma rays to fundamental physical processes occuring in SNe Ia permits
  additional constraints
  concerning several explosion model properties that are not easily
  accessible within other wavelength ranges. Proposed
  future MeV missions such as GRIPS will resolve all spectral details
  only for nearby SNe~Ia, but hardness ratio and light curve measurements 
  still allow for a distinction of the two different models at $10$
  and $16\,\mathrm{Mpc}$ for an exposure time of $10^6\,\mathrm{s}$, 
  respectively. The possibility to detect the
  strongest line features up to the Virgo distance will offer the 
  opportunity to build up a first sample of SN~Ia detections in the 
  gamma-ray energy range and underlines the importance of future space 
  observatories for MeV gamma rays.}
{}


\maketitle
%

\section{Introduction}
While there is general agreement that SNe Ia are the result of
thermonuclear explosions of carbon-oxygen WDs, many questions
concerning the progenitor and explosion scenarios still remain open
\citep[cf.][]{Hillebrandt2000}. This lack of knowledge is contrasted
by the relevance of SNe~Ia for measuring cosmological distances
as well as their influence on the chemical evolution of the universe
and emphasizes the need of a more thorough understanding of the
underlying physics. To constrain current explosion models as tightly 
as possible, a multi-wavelength approach extending to the gamma-ray 
regime (MeV energies) is desirable.

The gamma-ray spectra of SNe~Ia are dominated by the lines of
the $^{56}\mathrm{Ni} \rightarrow {^{56}\mathrm{Co}} \rightarrow
{^{56}\mathrm{Fe}}$ decay chain which powers the observable display
of SNe~Ia. In contrast to studies in the optical or infrared 
wavelength regime where the emissivity is strongly dependent on
the complex opacity structure which in turn depends on the atomic
level populations and chemical composition, the emissivities in the
gamma-ray regime are determined from their branching ratios 
and radioactive half-lives and a few rather simple interaction
processes like pair-production, Compton scattering and photo-electrical
absorption. This makes the gamma-ray regime an ideal tool to obtain a 
more direct handle on the mass-velocity distribution of the
explosion product \citep{Milne2004}. These very promising prospects 
have led to
numerous theoretical efforts (mostly one-dimensional) in investigating 
the gamma-ray emission
of SNe Ia in the past \citep[see for
instance][]{Clayton1969,Clayton1974,Ambwani1988,Chan1988,Chan1990,Chan1991,
  Burrows1990,Mueller1991,Hoeflich1992,Kumagai1997,Gomez-Gomar1998,Hoeflich1998,
  Sim2008,Isern2008,Kromer2010,Maeda2012}.
Although there have been first attempts of taking three-dimensional
effects into account \citep[e.g.][]{Hoeflich2002}, a fully three-dimensional
treatment of the explosion hydrodynamics as well as the radiation transfer
calculations has not been carried out in previous SN Ia gamma-ray emission
studies. We determine, to our knowledge for the first time, the predicted
gamma-ray emission for completely three-dimensional SN Ia explosion models. The
three-dimensional approach is important for a realistic description of the
distribution of the radioactive isotopes and the surrounding ejecta material --
both factors the gamma-ray emission of SNe Ia is sensitive for as shown
in the following. Furthermore, a thorough investigation of line-of-sight effects
due to different viewing angles relies on a three-dimensional treatment of the
explosion scenario.

However, as of now the detection of SNe~Ia at MeV energies has not been 
practicable owing to the low sensitivities or non-existence of detection
instruments in this energy range. In this paper, we explore the
additional benefits from the analysis of gamma-ray spectra towards a
more sound theoretical understanding of SNe Ia and discuss the 
detection limits of proposed next generation gamma-ray observatories. 
For this aim to be achieved, we run full detector simulations of the
proposed MeV satellite GRIPS \citep{Greiner2012}.

In this paper, we focus on two main branches of suggested SN~Ia progenitor
models: The explosion of a carbon-oxygen Chandra\-sekhar-mass WD and 
the super-Chandrasekhar-mass violent merger of two carbon-oxygen WDs. 
While the 
first scenario marks the end of a WD close to the Chandrasekhar limit 
that accretes mass from a companion star through Roche-lobe
overflow or strong stellar winds, the second 
is considered to be the result of two closely orbiting WDs losing
energy due to the emission of gravitational waves and
merging finally. 
The latter scenario has attracted a renewed interest in the last years
since, in contrast to previous thoughts, the publications by
\citet{Pakmor2010,Pakmor2011,Pakmor2012b} demonstrated that \textit{violent}
mergers of
two carbon-oxygen WDs can directly lead to a thermonuclear explosion while the
merger is still ongoing. As an extension to preceding gamma-ray studies of SNe
Ia, we discuss the gamma-ray emission of a violent merger for the first time.

Both scenarios generate optical observables that are
similar to those of normal SNe Ia
\citep{Mazzali2007,Kasen2009,Blondin2011,Pakmor2012a,Roepke2012}, in spite of
significant differences in the total mass and the ejecta structure.
Further observable distinctions through SNe Ia gamma-ray spectra would
therefore provide an additional handle on the progenitor channel
question.

The outline of the paper is as follows: In Sect.~\ref{sec:sim} we 
describe the applied explosion models and the radiative transfer scheme.
After a discussion of the resulting gamma-ray spectra and a study of the 
applicability of different diagnostic tools in Sect.~\ref{sec:res}, we 
investigate the chances of detecting gamma-ray line emission from SNe Ia 
in the near future in Sect.~\ref{sec:pros} and conclude with a short 
summary of the main findings.

\section{Simulation framework} \label{sec:sim}

\subsection{Explosion models}
In the first simulation, we calculate the gamma-ray emission for the
explosion of a near Chandrasekhar-mass (\mch) WD as a delayed
detonation \citep[e.g.][]{Khokhlov1991}. After an initial subsonic
deflagration phase of nuclear burning that
produces mainly iron group elements, the explosion turns into a 
supersonic detonation. The remaining fuel is mostly burned to
intermediate mass elements due to the prevailing lower densities
caused by the energy release of the deflagration mode and the
subsequent expansion of the star. We use the N100 model from the
set of three-dimensional delayed detonation simulations carried out by
\citet{Seitenzahl2013} with the thermonuclear supernova code
\textsc{Leafs}.  For a detailed description of the applied techniques
we refer the reader to \citet{Reinecke1999}, \citet{Roepke2005a}, 
\citet{Schmidt2006}, \citet{Roepke2007} and references
therein.  We chose the N100 model since it produces optical
observables that closely resemble those of ``normal'' SNe~Ia
\citep[][]{Roepke2012}.  The model is based on an isothermal
non-rotating WD in hydrostatic equilibrium with a central density of
$\rho_c = 2.9 \times 10^9\,\mathrm{g\,cm^{-3}}$.
During the explosion, $10^6$ tracer particles reproducing the underlying 
density profile of the WD record the thermodynamic conditions. We then use 
the information provided by the tracer particles to calculate the 
detailed isotopic composition in a post-processing step with a reaction 
network of 384 nuclides \citep{Travaglio2004,Roepke2006b,Seitenzahl2010}.
The initial chemical composition is assumed to be 47.5\,\% $^{12}\mathrm{C}$,
50.0\,\%
$^{16}\mathrm{O}$ and 2.5\,\% $^{22}\mathrm{Ne}$ by mass, resulting in
an electron fraction of $Y_e = 0.498864$, which corresponds to a 
zero-age main sequence metallicity comparable to that of the Sun.
With a kinetic energy of $1.45 \times 10^{51}
\mathrm{erg}$ and a total mass of $1.40\, \msun$ of the ejecta, N100 produced
$0.604\, \msun$ of \nuc{56}{Ni}.  Roughly half of the \nuc{56}{Ni} is
located in the inner $0.3\, \msun$ at velocities below $4,000\, \kms$,
while the other half is more or less isotropically but inhomogeneously
distributed within the remainder of the inner ${\sim}1.2\, \msun$, out
to velocities of $12,000\, \kms$.  

The second simulation is the violent merger
of a $1.1\,\msun$ and a $0.9\,\msun$ WD of \citet{Pakmor2012a}. Again, 
this model reproduces the features of ``normal'' SNe~Ia at optical
wavelengths \citep{Roepke2012}. To model the inspiral and the merger,
\citet{Pakmor2012b} used a modified version of the SPH code
\textsc{Gadget} \citep{Springel2005}.  Both WDs are constructed from a
total of $1.8\times10^6$ particles of equal mass. After a relaxation
phase, the distance between the WDs is slowly decreased according to the method
of
\citet{Dan2011} until the first particle of the less massive
(secondary) WD reaches the Lagrangian point between the two
objects. This triggers the actual simulation to start. During its
progress, more and more material from the secondary WD is accreted and
heated up on the surface of the primary, giving rise to the
formation of hot spots and the ignition of carbon burning. Following
the guidelines of microscopic detonation simulations
\citep{Seitenzahl2009}, a detonation is initiated in a hot spot that
reaches a temperature of more than $2.5\times10^9\,\mathrm{K}$ and a
density of about $2\times10^6\,\mathrm{g\,cm^{-3}}$. The mapping of
the actual simulation to a uniform Cartesian grid with
$768\times768\times768$ cells and a total box size of
$4\times10^9\,\mathrm{cm}$ is then used as initial state for a
simulation of the detonation flame with the \textsc{Leafs} code, where
the detonation is ignited at the cell with the highest temperature.
For more information about the simulation see
\citet{Pakmor2012b,Pakmor2012a}.
The detailed composition of the material is again calculated with the
tracer particle method and a post-processing step using a large
nuclear reaction network (see above).  The merger model yields are
based on the same initial chemical composition as the
delayed-detonation model (47.5\,\% $^{12}\mathrm{C}$, 50.0\,\%
$^{16}\mathrm{O}$ and 2.5\,\% $^{22}\mathrm{Ne}$ by mass).  With an
asymptotic kinetic explosion energy of $1.7 \times 10^{51}
\mathrm{erg}$, the merger model produced $0.616\, \msun$ of
\nuc{56}{Ni} out of the combined initial mass of $2.0\, \msun$. 
In contrast to the delayed-detonation model N100, the
\nuc{56}{Ni} is mainly found at velocities below
${\sim}10,000\,\kms$ and it is much more asymmetrically
distributed in the ejecta. This is mainly due to the delayed explosion
of the secondary into the already burned remains of the primary, which
excavates a region virtually free of iron group elements at low
velocity \citep[see figure~2 of][]{Pakmor2012b}.

\subsection{Radiative transfer}
Using detailed abundance distributions obtained from the tracer
particle method we map the explosion ejecta to a
$50\times50\times50$ Cartesian grid and follow the emission,
propagation and interaction of the gamma-ray photons 
with the Monte Carlo radiative transfer code \textsc{Artis}
\citep{Sim2007,Kromer2009}. The main principles of the
radiative transport code can be summarized as follows: In contrast to
Nature's way of quantization, the radiation field is divided into
Monte Carlo quanta representing indivisible parcels of energy,
providing several advantages concerning the simulation technique
\citep{Lucy1999}.  Initially, the quanta begin as so-called pellets of
radioactive material representing $^{56}\mathrm{Ni}$, 
${^{56}\mathrm{Co}}$, ${^{52}\mathrm{Fe}}$ and ${^{48}\mathrm{Cr}}$
(other radionuclei are neglected in \textsc{Artis} since they
are typically not important at early times). Upon decay, the pellets
are converted to mono-chromatic gamma-ray packets with frequencies 
sampled randomly according to the
respective branching ratios. These gamma-ray packets are emitted into
randomly chosen directions under the assumption of isotropic emission 
in the comoving frame. Then, their
propagation is followed in frequency, three-dimensional space, and
time, until they leave the ejecta or are removed from the gamma-ray
regime due to interaction processes. The basic interaction processes 
of gamma-ray 
photons are pair production, photoelectrical absorption, and Compton
scattering, with the latter being the most dominant in the
encountered energy ranges \citep{Milne2004}. In accordance with the ratios of
the cross sections of individual interaction processes 
to the total cross section, the occurrence of
a certain interaction type is sampled randomly. Throughout the
simulation, a positronium fraction of zero is assumed, meaning that
positrons, e.g.\ from pair production or nuclear decays, annihilate in
situ and directly lead to the production of two gamma-ray photons at
$0.511\,\mathrm{MeV}$ \citep[cf.][]{Milne2004}. The escaping gamma-ray
packets are binned in frequency, time and direction, contributing to
the spectral evolution of the gamma-ray emission from the explosion
event. Light-travel time effects are taken into consideration. A
thorough description of the employed Monte Carlo radiative transfer
scheme and additional references can be found in
\citet{Lucy2005}, \citet{Sim2007}, \citet{Sim2008} and \citet{Kromer2009}.

\begin{figure}[!]
  \centering
  \includegraphics[width=8.93cm]{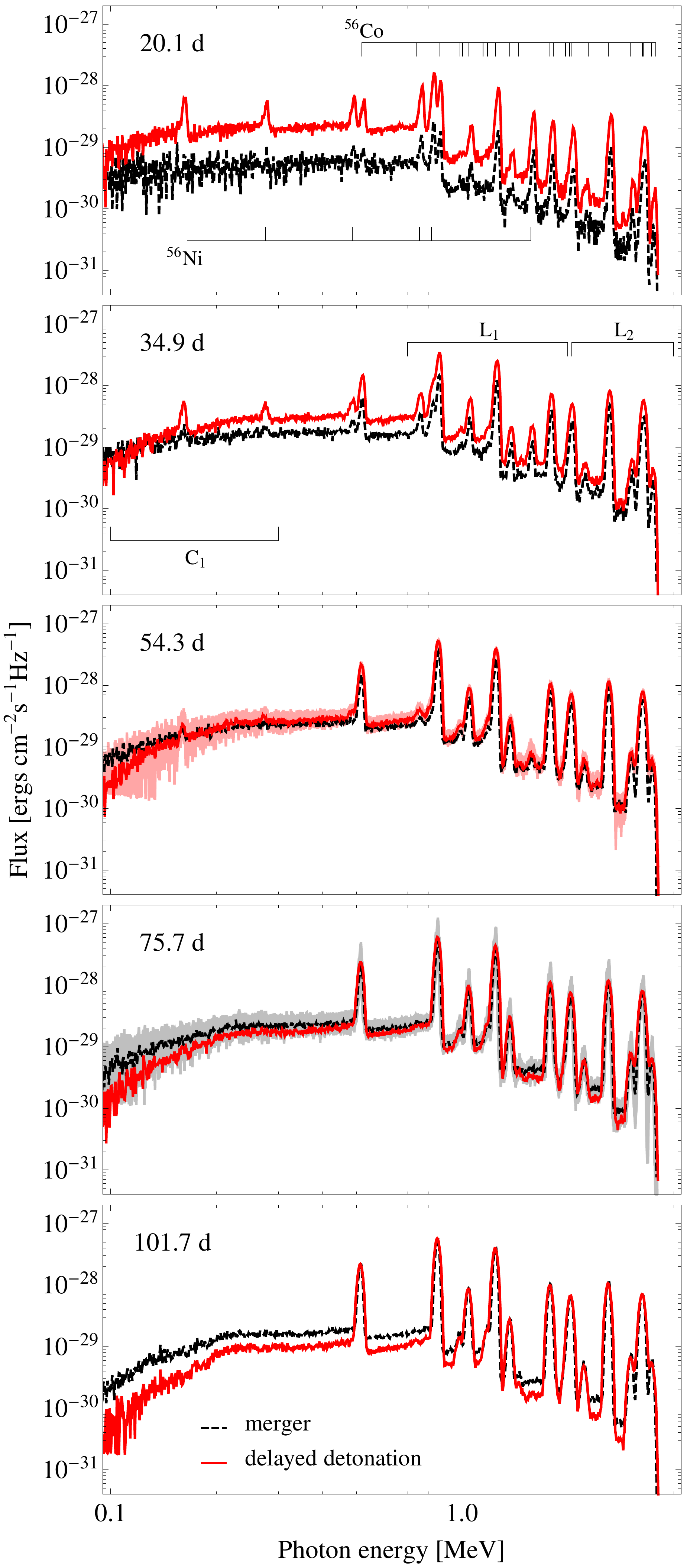}
  \caption{Spectral evolution of the gamma-ray emission from
    the delayed detonation (red) and the violent merger model (dashed black)
    for different epochs after the explosion (indicated in the upper 
    left of each panel). The spectral spread due to different viewing angles is
    shown for the maximum-light epochs of the two models in
    gamma rays (indicated in light red for the delayed detonation model in the 
    third panel and in gray for the merger model in the fourth panel). 
    Part of the effect is obscured by Monte Carlo noise in particular in the 
    continuum. This problem is largely removed by using hardness ratios and 
    broadband light curves. The $^{56}\mathrm{Ni}$ 
    and $^{56}\mathrm{Co}$ emission lines are indicated in the first panel, 
    the flux bands $C_1$, $L_1$ and $L_2$ 
    discussed in Sect.~\ref{line_hard_rat} are 
    depicted in the second panel.}
  \label{spec_evol}
\end{figure}

\section{Results and discussion} \label{sec:res}
Even though the Chandrasekhar-mass delayed detonation and the violent
merger model show distinct differences in total mass and ejecta
structure, both models produce nearly the same amount of
$^{56}\mathrm{Ni}$ (roughly $0.6\, \msun$). In spite of their
differences, a straightforward preference to one of the models 
cannot be given by comparing the simulated optical spectra to the
measured for the test case of SN 2011fe \citep{Roepke2012}. 
In this section, we investigate whether
additional discriminating features can be identified from a study of
the gamma-ray emission of the two models.
 
\subsection{Gamma-ray spectra} \label{sec:spectra}
Fig.~\ref{spec_evol} shows the spectral evolution of the
angle-averaged gam\-ma-ray emission arising from the delayed detonation
and the violent merger model (photon fluxes are always normalized to a
distance of $1\,\mathrm{Mpc}$). For a discussion of the importance of
viewing angle effects, the reader is referred to the end of this
section. 

The spectra are dominated by lines of the decay chain
${^{56}\mathrm{Ni}}\rightarrow{^{56}\mathrm{Co}}\rightarrow{^{56}\mathrm{Fe}}$.
Due to the different half-lives of $^{56}\mathrm{Ni}$ $(6.1\,\mathrm{d})$
and $^{56}\mathrm{Co}$ $(77.2\,\mathrm{d})$, $^{56}\mathrm{Ni}$ lines
which dominate at early epochs (e.g.\ at $0.812\,\mathrm{MeV}$ or
$1.562\,\mathrm{MeV}$) vanish in the spectra at later times, according
to the decreasing amounts of $^{56}\mathrm{Ni}$. Then, the spectra
are mainly formed by strong emission lines of $^{56}\mathrm{Co}$ and a
continuum contribution due to Compton scattering of line
photons. The optical depth to Compton scattering mainly depends on the 
column density of target electrons. Since the energies of 
gamma-ray photons of ${\sim}1\,\mathrm{MeV}$ are
much higher than the corresponding binding energies of electrons in
atoms, nearly all electrons, regardless if bound or unbound, are 
accessible as targets of Compton scattering processes.
As the ejecta expand with time, optical depths are reduced, leading 
to an enhancement of the lines with respect to the continuum.

\begin{figure}
  \centering
  \includegraphics[width=9cm]{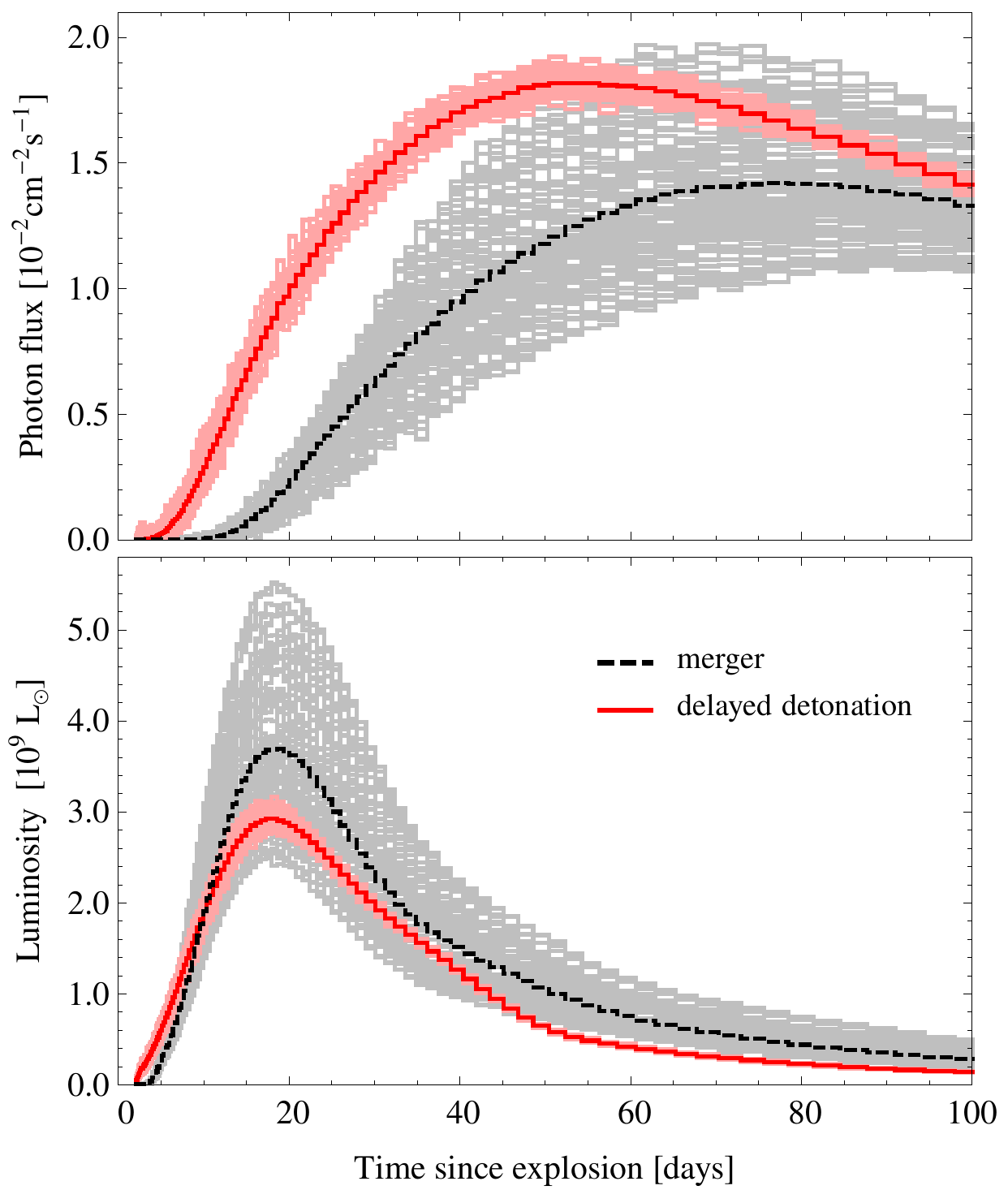}
  \caption{Bolometric gamma-ray light curve (upper panel) and bolometric 
    UVOIR light curve (lower panel) for the delayed detonation
    (red) and the violent merger model (dashed black). The light curve spread
    due to different viewing angles is indicated in light red and
    gray. The photon fluxes resp.\ the luminosities are normalized to a 
    distance of $1\,\mathrm{Mpc}$.}
  \label{light_bol}
\end{figure}

Besides these common properties of both explosion models, there are
also some pronounced differences. The gamma-ray emission in the
delayed detonation scenario evolves faster than in the WD-WD merger, a
fact that is also visible in a comparison of the bolometric gamma-ray 
light curves (energy range from $0.05$ to $4.0\,\mathrm{MeV}$) of the
two models (see Fig.~\ref{light_bol}, upper panel). The delayed detonation 
produces a peak photon flux of
$1.82\times10^{-2}\,\mathrm{cm^2\,s^{-1}}$ at $54.3\,\mathrm{days}$, 
whereas the merger yields a maximum
flux of $1.43\times10^{-2}\,\mathrm{cm^2\,s^{-1}}$ at
$75.7\,\mathrm{days}$ after the explosion. Since the masses of
produced radioactive $^{56}\mathrm{Ni}$ are nearly equal in both
models, this is purely a consequence of the fact that the transport of
gamma rays is, to first order, only sensitive to the column density of
the material above the emission region.  I.e.\ the larger total
mass of the WD-WD merger delays the gamma-ray emission and gives rise
to a lower (angle-averaged) peak flux compared to the delayed
detonation model. The convergence of the two light curves at late times
in Fig.~\ref{light_bol} reflects the similarity of the $^{56}\mathrm{Ni}$
mass in both models. In the optically thin limit, the gamma-ray
luminosity is given by
\begin{equation}
  L_\gamma(t\ge t_{\mathrm{thin}})\approx1.23\times10^{43}\frac{M_{\mathrm{Ni  
}}}{\msun}\exp\left(-\frac{t}{t_{\mathrm{Co}}}\right)\,\mathrm{erg\,s^{-1}}
\end{equation}
where $t_{\mathrm{Co}}$ is the lifetime of $^{56}\mathrm{Co}$
\citep{Sim2008}.  Thus, if the distance to the object is known, late-time
measurements of gamma-ray luminosities can be used to unambiguously
determine the explosion yield of $^{56}\mathrm{Ni}$.

Another distinguishing feature can be seen in the early-time spectra of the
delayed detonation model at $20.1$ and $34.9\,\mathrm{days}$
after the explosion: While two prominent lines of
$^{56}\mathrm{Ni}$ can be identified at $0.158\,\mathrm{MeV}$ and
$0.270\,\mathrm{MeV}$, these two lines are nearly totally degraded in
the merger model and vanish in the background of continuum
emission. This effect can be explained by the energy sensitivity of
the Compton cross section. Since the cross section decreases
with increasing photon energy, especially lines at low energies
suffer from effective Compton down-scattering and the additional
contamination of likewise down-scattered higher energy photons. Therefore, the
two
lines can only build up if a significant amount of $^{56}\mathrm{Ni}$
is located at small optical depths \citep[cf.][]{Gomez-Gomar1998}, a
fact that directly connects the occurrence of low-energy
$^{56}\mathrm{Ni}$ lines to the distribution of the radioactive
material. These different distributions of $^{56}\mathrm{Ni}$
can be seen in figures 1 and 2 of 
\citet{Roepke2012}, where it is shown that there is much more
$^{56}\mathrm{Ni}$ at higher velocities in the delayed detonation
model than in the merger model. Furthermore, there is less material 
surrounding the $^{56}\mathrm{Ni}$ regions in the delayed detonation
model and the corresponding column densities are therefore lower than 
in the merger scenario. This property is clearly mirrored in
the evolution of the gamma-ray emission, but cannot be inferred
easily from measurements in other wavelength ranges (see e.g.\ figure 3
in \citealt{Roepke2012}). The larger optical depths
outside the $^{56}\mathrm{Ni}$ region lead to more efficient
Compton down-scattering and thus a softer spectrum of the
merger model. This is further enhanced by the fact that iron-group 
elements are confined to lower velocities in the merger, leading to less
photoelectric absorption than in the delayed detonation.

\begin{figure*}
  \centering
  \includegraphics[width=18cm]{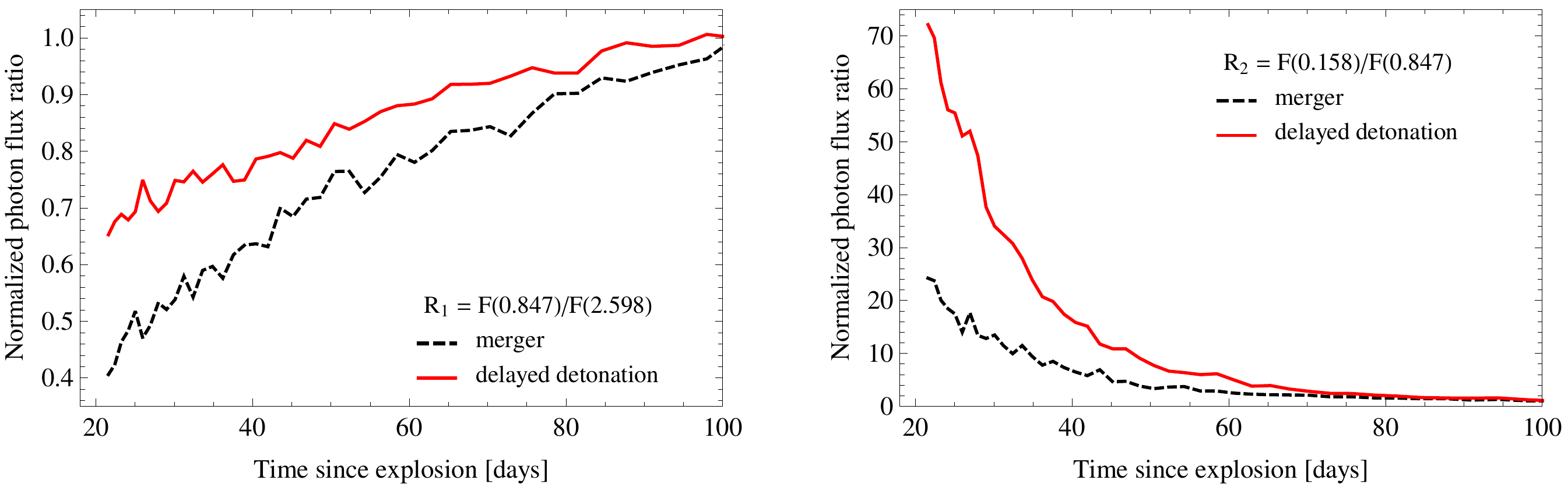}
  \caption{Peak line flux ratios of the gamma-ray emission from the
    delayed detonation (red) and the violent merger model (dashed black). 
    The graph on the left depicts the line ratio
    $R_1=F(0.847\,\mathrm{MeV})/F(2.598\,\mathrm{MeV})$ of two
    $^{56}\mathrm{Co}$ lines, the graph on the right illustrates the
    line ratio $R_2=F(0.158\,\mathrm{MeV})/F(0.847\,\mathrm{MeV})$ of
    a $^{56}\mathrm{Ni}$ and a $^{56}\mathrm{Co}$ line. The flux
    ratios are normalized to the optically thin limit.}
  \label{line_rat}
\end{figure*}

\begin{figure*}
  \centering
  \includegraphics[width=18cm]{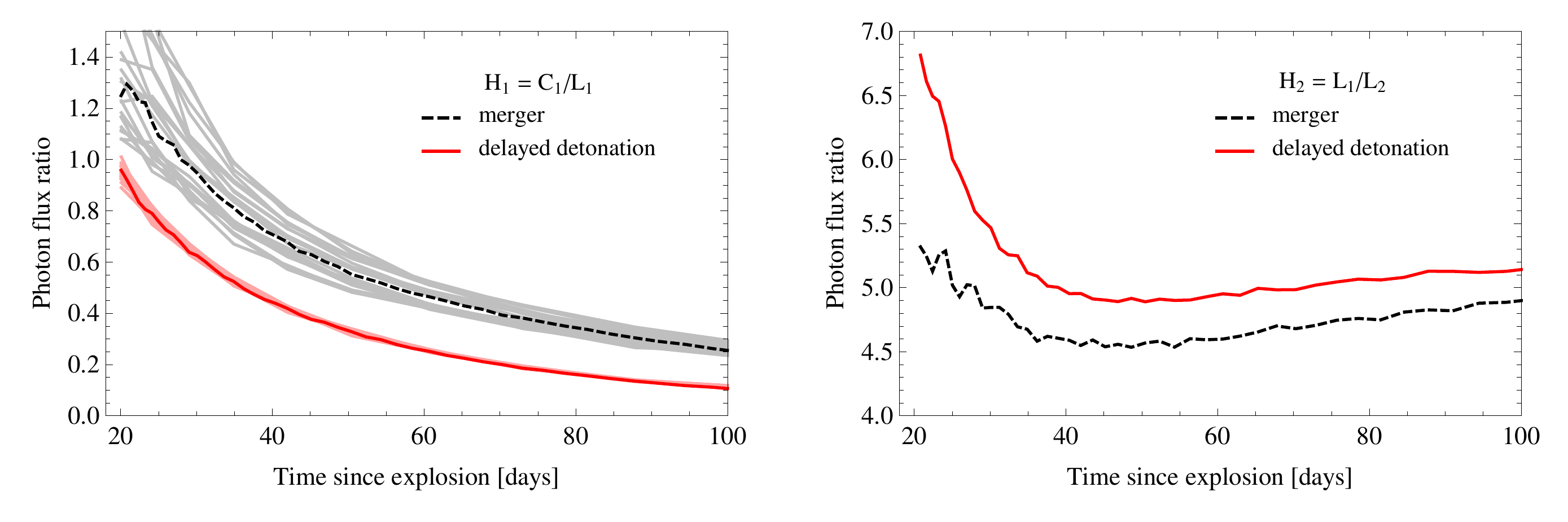}
  \caption{Hardness ratios of the gamma-ray emission from the delayed
    detonation (red) and the violent merger model (dashed black). The ratio
    $H_1=C_1/L_1$ is shown on the left, the ratio $H_2=L_1/L_2$ is
    displayed on the right. To illustrate the sensitivity to viewing-angle 
    effects, $H_1$ is shown for different lines of sight in the left panel.}
  \label{hard_rat}
\end{figure*}

The influence of different viewing angles on the gamma-ray spectra is
illustrated in Fig.~\ref{spec_evol} for the maximum-light epochs 
of the two models in gamma rays. While the
strongest lines of the delayed detonation model do not show much
variation, the asymmetric structure of the ejecta in the merger model
leads to spectral features of varying magnitude, and for certain
viewing angles a distinction from the spectra of the delayed
detonation model can be very difficult. The low-energy range of the
spectra is still the most suitable for distinguishing the two models:
In the merger model, the higher column densities due to the larger 
total mass result in more effective Compton down-scattering of the 
gamma-ray photons in this energy range than in the delayed detonation model. 
This hinders the formation of prominent low-energy $^{56}\mathrm{Ni}$ lines. The
effects of observing the two explosion models at different lines of
sight are also illustrated for the bolometric gamma-ray light curves
in Fig.~\ref{light_bol}. As before, the asymmetric distribution of ejecta
in the merger model results in a much larger spread of photon
fluxes compared to the delayed detonation model. Nevertheless, the different
times of maximum flux, as well as the differing early time evolution of 
the light curves, provide clear distinctive features that can be used to
discriminate the underlying explosion models. These differences are
again due to the different total masses involved in the two
explosion scenarios. The larger mass of the merger model leads to more
efficient photon trapping at earlier times, delays the rise of the
photon flux, and results in a flux peak occurring at later times.

As a comparison, we show the bolometric UVOIR light curves of the two models 
in the lower panel of Fig.~\ref{light_bol}. For many viewing 
angles, a distinction between the delayed detonation and the merger model is
nearly impossible. This is contrasted by the behavior of the bolometric 
gamma-ray light curves: Here, in spite of viewing angle effects, 
especially early time measurements as well as the determination 
of the maximum fluxes are very promising. This again underlines 
the advantages of gamma-ray emission studies of SNe Ia. 

In the following, we discuss further possibilities to discriminate the
two explosion models on the basis of angle-averaged spectra and
present the fundamental distinguishing characteristics. As described
in the previous paragraph, viewing-angle effects can considerably
complicate the process of drawing inferences about certain explosion
scenarios, but distinctions are still possible.

\subsection{Line and hardness ratios} \label{line_hard_rat}
In addition to the analysis of line fluxes and light curves, line and
hardness ratios represent further diagnostic tools that can be
utilized for the study of the gamma-ray emission of SNe~Ia 
\citep[cf.][]{Hoeflich1998,Gomez-Gomar1998,Sim2008}. While the
information inferred from absolute flux values is always restricted by
how well the distance to the source is known, flux ratios are 
distance-independent and not subject to this limitation. In
Fig.~\ref{line_rat}, two characteristic peak-intensity line ratios of
the WD-WD merger and the delayed detonation model are illustrated.
Following \citet{Sim2008} we define 
$R_1=F(0.847\,\mathrm{MeV})/F(2.598\,\mathrm{MeV})$ as the 
ratio of
two $^{56}\mathrm{Co}$ lines
and $R_2=F(0.158\,\mathrm{MeV})/F(0.847\,\mathrm{MeV})$ as the ratio
of a $^{56}\mathrm{Ni}$ and a $^{56}\mathrm{Co}$ line.

Flux ratios are affected by similar processes as discussed in the
previous paragraph: Due to the energy sensitivity of Compton cross
sections, line ratios are, at early times before the optically thin
limit, dependent on the column density of electrons and therefore on the
material above the region containing the radioactive species. The
ratio of two lines of the same radioactive isotope is therefore
determined by the opacity
ratios at the respective line energies. This is shown in the left plot
of Fig.~\ref{line_rat} for the case of $R_1$. Since the radioactive
material in the merger model is behind much more opacity, the 
line ratio shortly after the explosion is much lower. This is different 
in the delayed detonation model: Here, more
$^{56}\mathrm{Ni}$ and hence more $^{56}\mathrm{Co}$ at lower optical
depths increases the line ratio significantly. It is thereby 
not crucial to choose two specific $^{56}\mathrm{Co}$ lines for the
examination of such an effect, but the extent of the diagnostic
validity relies, due to the energy dependence of the Compton cross
section, on a sufficient spread between the energy of the two 
selected lines \citep[cf.][]{Sim2008}.

In contrast to $R_1$, $R_2$ is a function that decreases with
time. Since $R_2$ is a ratio of two lines from different isotopes,
this behavior reflects the difference between the half-lives of
$^{56}\mathrm{Ni}$ and $^{56}\mathrm{Co}$: While the
$^{56}\mathrm{Co}$ line at $0.847\,\mathrm{MeV}$ strengthens, the
$0.158\,\mathrm{MeV}$ line of $^{56}\mathrm{Ni}$ fades away at later
times, leading to smaller peak flux ratios. The information value of
$R_2$ is based on the following aspect: As discussed before, 
the $^{56}\mathrm{Ni}$ present in the outer shells at small optical depths
is the main source of the emerging $0.158\,\mathrm{MeV}$ line,
whereas the emission of the $0.847\,\mathrm{MeV}$ line originates from the
total abundance of $^{56}\mathrm{Co}$ in the ejected material,
especially at later times. Therefore, a larger value of $R_2$ indicates a
larger 
deposit of $^{56}\mathrm{Ni}$ in the outer layers of the ejecta and the 
different distributions of this isotope in the two models are clearly reflected 
in the evolution of the line ratios shown in Fig.~\ref{line_rat}. As stated
before
\citep{Gomez-Gomar1998,Sim2008}, the relatively low energy of the
$0.158\,\mathrm{MeV}$ line makes it also very sensitive to
photoelectric absorption processes. Since photoelectric opacities are 
dependent on the compositional structure of the ejecta above the 
radioactive material, appropriate ratios of low-energy
$^{56}\mathrm{Ni}$ lines to $^{56}\mathrm{Co}$ lines at higher
energies are well suited to study the composition of SNe~Ia.
The success in finding distinct abundance features of course
relies on the quality of the measured gamma-ray data and for this
reason also on the distance to the explosion site (see also the
remarks in the next section).

A way to partially circumvent the need for highest quality gamma-ray
data for detailed SNe~Ia studies is provided by the application of
hardness ratios.  Instead of discrete line intensities, the fluxes of
broader energy bands are compared to each other. On the one hand, in this
coarser method, some information is necessarily lost.
But on the other hand, in contrast to other wavelength ranges, the 
relative simplicity of gamma-ray spectra and the rather small number of factors
that influence gamma-ray emission processes make
hardness ratios an important diagnostic tool. Following
\citet{Sim2008}, Fig.~\ref{hard_rat} shows two hardness ratios, 
$H_1=C_1/L_1$ and $H_2=L_1/L_2$, as examples for the 
delayed detonation and the merger model. $C_1$ denotes the energy 
band from $0.1$ to $0.3\,\mathrm{MeV}$, where the main contribution
comes from continuum emission by Compton down-scattering and
photo-absorption processes. The energy bands $L_1$ (from $0.7$ to
$2.0\,\mathrm{MeV}$) and $L_2$ (from $2.0$ to $4.0\,\mathrm{MeV}$) are
characterized by pronounced lines of $^{56}\mathrm{Co}$, while the
importance of continuum emission decreases with higher photon
energies.

Similar to line ratios, hardness ratios are mainly affected by the energy
dependence of the Compton cross section. Since $H_1$ directly mirrors
the strength of the continuum emission relative to that of the lines,
the violent merger produces higher $H_1$ values than the delayed
detonation -- a direct consequence of the radioactive material's
location at higher optical depths. Similar to the line ratio $R_2$,
$H_1$ is an indicator of different column densities of 
target electrons. Since photoabsorption processes can be significant in 
the energy band $C_1$, $H_1$ also serves to a certain degree as 
an indicator of different compositions. $H_2$
represents the flux ratio of two line-dominated energy bands and is
analogous to the line ratio $R_1$. It depends mainly on the ratio of
the Compton cross sections in the two corresponding energy bands and
is also suitable to discriminate our explosion models. Hence, hardness
ratios offer an alternative opportunity to extract information from
the gamma-ray emission of SNe~Ia -- particularly for more
distant explosion events.

The results in Fig.~\ref{line_rat} and \ref{hard_rat} show that line
and hardness ratios can be used as additional diagnostic tools to 
distinguish the delayed detonation from the merger model on the basis 
of their gamma-ray emission. These ratios
are directly linked to the distribution of $^{56}\mathrm{Ni}$ in the 
ejecta and, due to their dependence on the column 
density of target electrons above the radioactive material, they are 
sensitive to different compositions and masses of the outer ejecta layers. 
Even in light of possible viewing angle effects (see left panel of
Fig.~\ref{hard_rat} as an example), hardness ratios have proven to be 
quite robust in distinguishing our two models. Together 
with broadband light curve measurements, they provide the best chances for 
conclusions on certain model features especially for more distant explosion 
events.  

However, besides the changes in the gamma-ray emission due to different
viewing angles, other variations within a certain explosion scenario (e.g.
additional rotation and mixing effects, modifications of details in the
explosion mechanism) can lead to a spread in gamma-ray fluxes that makes it
certainly very difficult to distinguish between different realizations of SN Ia
explosions. But concerning our two very different models representing two
different progenitor classes, we are confident that the delayed detonation and
the violent merger as described above will leave a unique imprint on the 
gamma-ray emission: In the case of the same $^{56}\mathrm{Ni}$ mass in both models, 
a higher total ejecta mass surrounding the radioactive material in the applied 
violent merger is unavoidable. Corresponding to
our results that we have discussed before, this has a systematic effect on the
gamma-ray observables. As we will show in the next section, the prospects for
discriminating our two models in case of a nearby SN Ia are very promising, but
rely on the realization of new MeV satellite missions in the near future.

\section{Detection prospects} \label{sec:pros}
Despite the importance of the MeV energy range for the study of
different astrophysical objects and processes
\citep[cf.][]{Greiner2012,Summa2011}, the mission
with the highest sensitivities at these energies was the COMPTEL 
instrument aboard the CGRO
satellite in the mid-nineties. One possible successor with an enhanced
sensitivity of a factor 40 is the proposed GRIPS (Gamma-Ray Imaging, 
Polarimetry and Spectroscopy) mission
\citep{Greiner2009,Greiner2012}. The focus of this section is on the
progress that could be made in the future concerning the detection of 
gamma-ray emission from SNe~Ia, using the GRIPS instrument as an example.

\begin{figure}
  \centering
  \includegraphics[width=9cm]{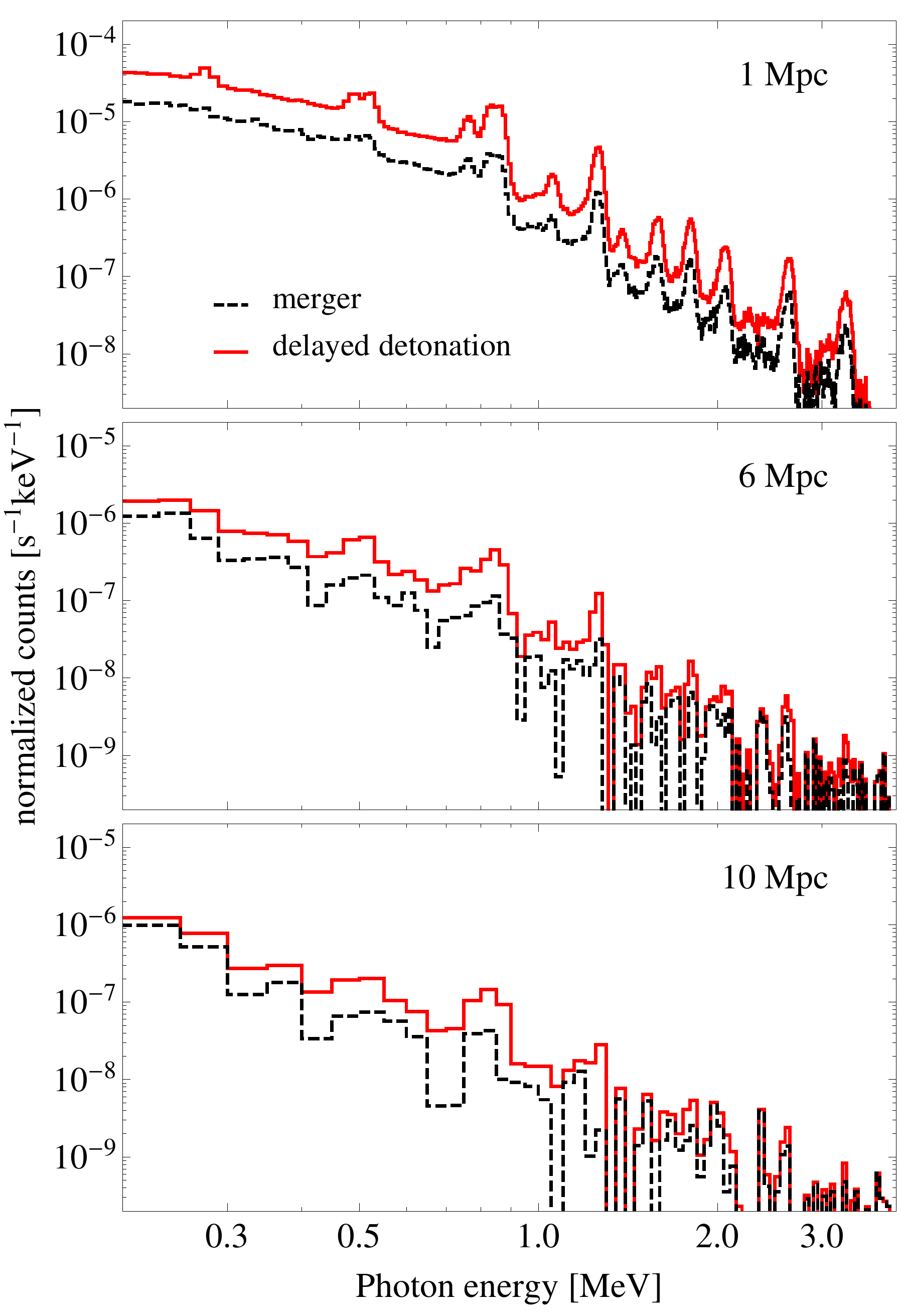}
  \caption{Simulated detector response for the GRM instrument aboard
    GRIPS using an exposure time of $10^5\,\mathrm{s}$. Shown is 
    a comparison of the simulated angle-averaged gamma-ray spectra of the
    delayed detonation model (red) and the merger model (dashed black) at
    $25\,\mathrm{days}$ after the explosion for different distances to
    the source (indicated in the upper right).}
  \label{grips2}
\end{figure}

Being able to detect photons with energies from $0.2\,\mathrm{MeV}$ to
$80\,\mathrm{MeV}$, GRIPS is specifically designed for the search of
gamma-ray bursts and blazars, the study of particle acceleration and
radiation processes in a variety of cosmic sources, and the
study of supernova explosion and nucleosynthesis
mechanisms. The main instrument of GRIPS is the Gamma-Ray Monitor
(GRM), a combined Compton scattering and pair creation telescope
consisting of two separate detectors with an effective area of 
$195\,\mathrm{cm^2}$ and an energy resolution of $17\,\mathrm{keV}$ 
at $1.8\,\mathrm{MeV}$. The so-called \emph{tracker}, made of silicon
strips, is the first detector where the initial interaction of the
incoming gamma-ray photons takes place. Except for the entrance
surface it is surrounded by the second detector, a calorimeter
composed of $\mathrm{LaBr}_3$ scintillator material that allows for
the energy determination of the secondary particles. In the case of a Compton
scattering event in the tracker, the incident gamma-ray photon
interacts with an electron, whose energy and position can be
measured. The scattered photon is recorded in the calorimeter where
its energy and interaction point can be reconstructed. With these
data, it is possible to calculate the direction and the energy of the
incident photon. In the case of a pair creation event, the incident
gamma-ray photon is converted into an electron-positron pair within
the first detector. The original direction of the incident photon can
be determined from the tracked directions of these two particles. The
energy of the two secondaries, and therefore the energy of the incident
photon, is measured with the help of both the tracker and the
calorimeter.

\begin{figure*}
  \centering
  \includegraphics[width=18cm]{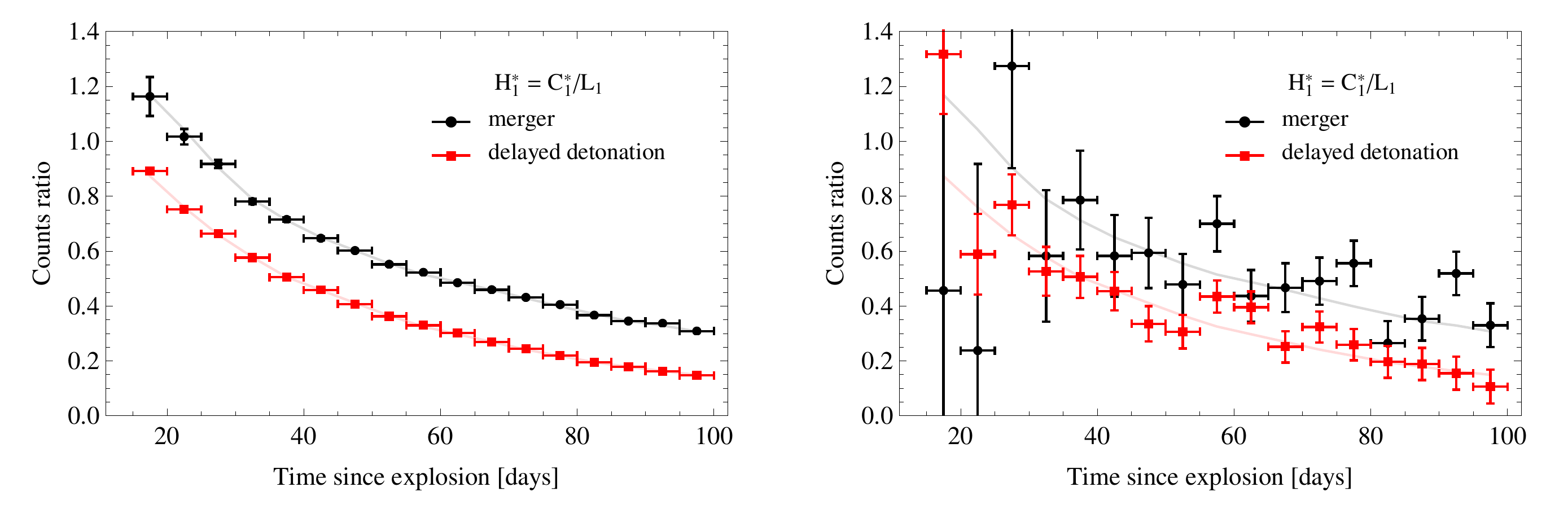}
  \caption{Simulation results for hardness ratio measurements with the
    GRM instrument (exposure time $10^5\,\mathrm{s}$) for our two
    explosion models. On the left, a source distance of
    $1\,\mathrm{Mpc}$ is assumed, the right figure shows the results
    for a source distance of $5\,\mathrm{Mpc}$. Taking into account
    the sensitivity limits of GRIPS, $C^*_1$ here denotes the energy
    band from $0.2$ to $0.4\,\mathrm{MeV}$. The solid lines show
    the results of ideal measurements without background fluctuations
    and statistical errors.}
  \label{hard_rat2}
\end{figure*}

\begin{figure*}
  \centering
  \includegraphics[width=18cm]{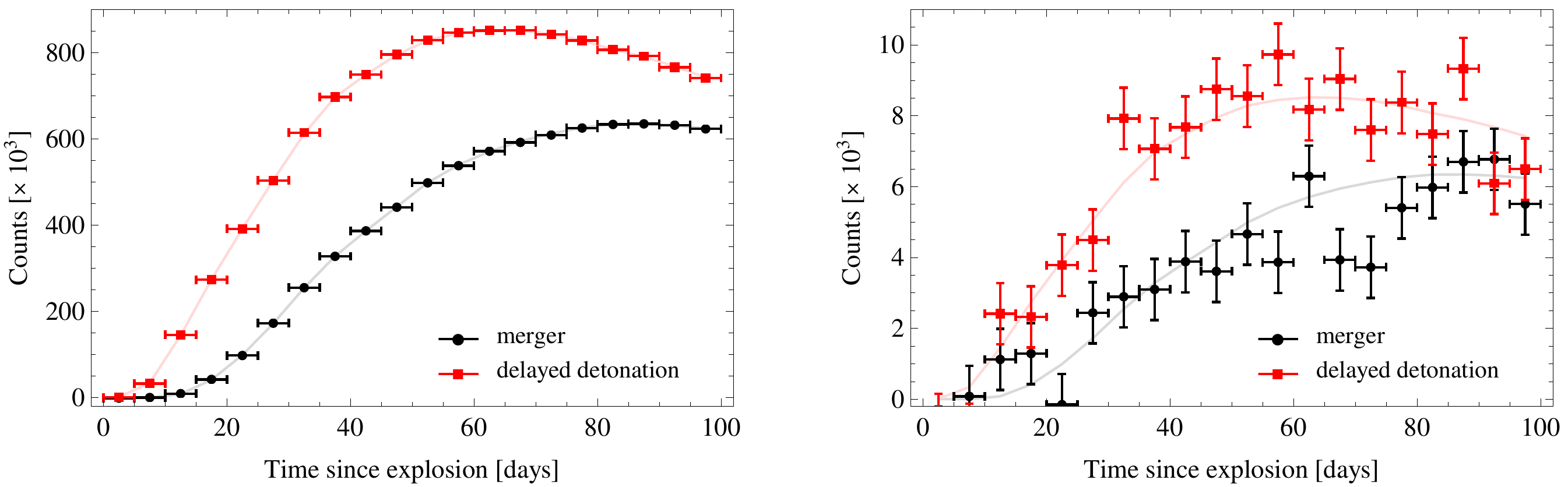}
  \caption{Simulation results for light curve measurements (energy
    range from $0.4$ to $4.0\,\mathrm{MeV}$) with the GRM instrument
    (exposure time $10^5\,\mathrm{s}$) for our two explosion
    models. On the left, a source distance of $1\,\mathrm{Mpc}$ is
    assumed, the right figure shows the results for a source distance
    of $10\,\mathrm{Mpc}$. The solid lines show the results of
    ideal measurements without background fluctuations and statistical
    errors.}
  \label{light_bol2}
\end{figure*} 

GRIPS measurements of gamma-ray emission from SNe~Ia were simulated
using the {\sc MEGAlib} software package \citep{Zoglauer2006}. This package
uses the {\sc Geant4} toolkit \citep{Agostinelli2003} to simulate the
passage of gamma rays and other particles through the detector, taking
into account possible particle interactions and decays. The detector
response to an incident gamma photon is calculated from the energies
deposited by the particles in sensitive elements (silicon strips and
scintillator crystals), taking into consideration instrumental effects
such as the energy resolution and noise suppression
thresholds. {\sc MEGAlib} includes a dedicated event reconstruction
algorithm, which is applied to the events that pass the trigger
criteria (coincident hits in the silicon strip detector and the
calorimeter). This algorithm either reconstructs the event as a valid
Compton scattering or pair creation event, or rejects it.  In the case
of a Compton event, the incident photon direction is constrained to a
circle around the direction of the scattered photon. To select the
events compatible with the source position of the photons, we required the
minimum
distance between the circle and the source position to be less than
$2^\circ$. In addition, if the recoil electron left a track in the
silicon detector, the Compton scattering plane could be reconstructed
and its rotation angle around the direction of the scattered photon
was required to be within $30^\circ$ from the source position.  The
model of the GRM detector used in this study corresponds to the setup
described in \citet{Greiner2012}.

Measurements of gamma-ray emission from most sources are strongly
affected by the presence of radiation backgrounds. In our simulations
we include the background from diffuse cosmic photons based on
\citet{Gruber1999} and the background from albedo photons produced in the
Earth's atmosphere \citep{Mizuno2004,Ajello2008}. In \citet{Boggs2006}
these two components were found to be the dominating background for a
Compton telescope at a low-inclination low Earth orbit. The generated
background photons were processed through the same simulation and
selection procedure described above and then the reconstructed
background events were added to the source events.

In order to account for the specific sensitivity range of GRIPS, we 
define here the hardness ratio $H^*_1$ 
analogous to $H_1$ as the quotient $C_1^*/L_1$, 
where $C_1^*$ denotes the energy band from $0.2$ to $0.4\,\mathrm{MeV}$. 
In Fig.~\ref{grips2}, the simulation results for the measured
gamma-ray spectra are shown for the two models at $25\,\mathrm{days}$
after the explosion. The exposure time of $10^5\,\mathrm{s}$
corresponds to roughly 5\,days in the all-sky scanning mode of GRM, a
time interval that allows for reasonable studies of the time
evolution. Accurate measurements of the line and hardness
ratios become difficult for large source distances due to background
fluctuations and limited event statistics in the high energy part of
the spectrum. For the GRIPS mission, nearly the best discrimination
between the two explosion models is provided by measurements of the
$H^*_1$ hardness ratio.
The simulation outcomes of the measurement of $H_1^*$ are depicted in 
Fig.~\ref{hard_rat2}. This ratio can be used to distinguish the models 
up to a source distance of $5$ to $7\,\mathrm{Mpc}$. Our simulation 
results for light curve measurements are shown in
Fig.~\ref{light_bol2}. For a supernova event at a distance of 1\,Mpc, a very
detailed and accurate light curve can be measured and constraints on
certain explosion models are possible. According to our analysis of
light curve measurements, GRIPS should be able to easily distinguish
the two explosion scenarios for a source distance up to $10\,\mathrm{Mpc}$.

\begin{figure}
  \centering
  \includegraphics[width=9cm]{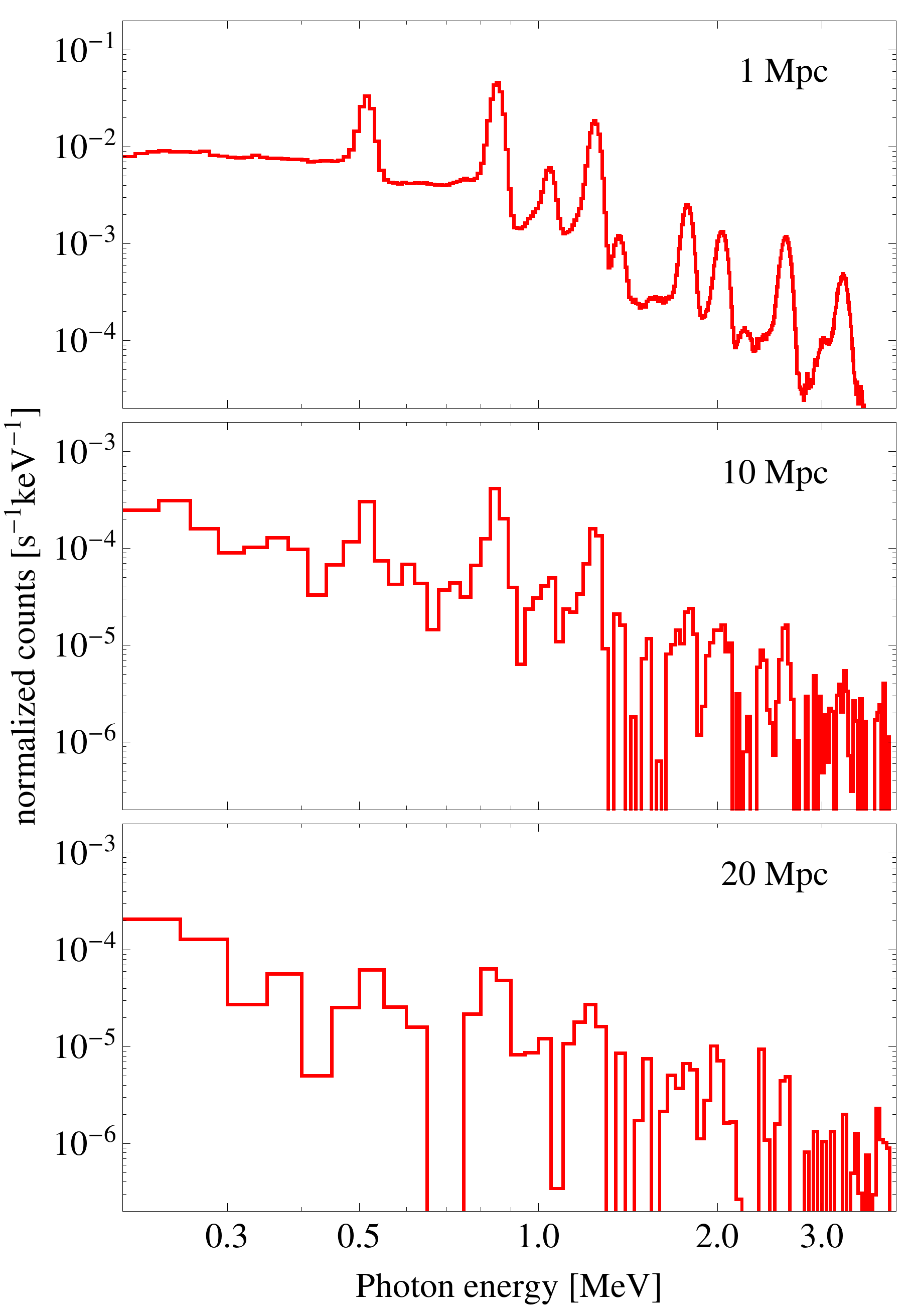}
  \caption{Simulated detector response for the GRM instrument aboard
    GRIPS for an exposure time of $10^6\,\mathrm{s}$. 
    Shown is the simulated angle-averaged gamma-ray 
    spectrum of the delayed detonation
    model at $60\,\mathrm{days}$ after the explosion for different
    source distances (indicated in the upper right).}
  \label{grips}
\end{figure}

Using the gamma-ray spectrum of the delayed detonation model at
$60\,\mathrm{days}$ after the explosion as input, we additionally
carried out detector response simulations for different source
distances and a longer exposure time of $10^6\,\mathrm{s}$ (roughly
$12\,\mathrm{days}$ in on-axis pointing mode).  The detection
capabilities of GRIPS concerning the gamma-ray emission of SNe Ia 
are illustrated in Fig.~\ref{grips}, where the simulated measurements
after background subtraction are depicted. While the strongest lines
at medium and higher energies can still be resolved even for sources
at $20\,\mathrm{Mpc}$ and more, the low energy range of the measured
spectra from sources at larger distances is dominated by background
fluctuations. Although GRIPS will provide a spectral resolution of all
details in the energy range from $0.2$ to $80\,\mathrm{MeV}$
only for SNe~Ia at distances up to a few Mpc, the application of line
and hardness ratios enables valuable spectral studies for even larger
distances. For an exposure time of $10^6\,\mathrm{s}$, our simulations 
show that the hardness ratio $H^*_1$ can be used to distinguish the two 
models up to $10\,\mathrm{Mpc}$. With this longer exposure time, GRIPS
light curve measurements can serve as distinctive marks of our models 
up to distances of $15$ to $16\,\mathrm{Mpc}$. Even 
explosion events in the Virgo cluster will be accessible for gamma-ray 
studies with such an instrument.

The success of such investigations of course depends on the accuracy
of the applied background models (see above). According to
\citet{Ajello2008}, an uncertainty of a factor of three seems to be
quite reasonable. Although the chances for a clear
discrimination between certain explosion models decrease considerably
for larger source distances, the ability of GRIPS in proving the
gamma-ray emission of SNe~Ia at distances beyond $20\,\mathrm{Mpc}$
will lead to a significant improvement of SNe~Ia detection statistics
in the gamma-ray energy range. Due to the low sensitivities of previous 
gamma-ray observatories, only one SN~Ia has been detected up to now 
\citep[cf.][]{Milne2004}.

In cases of events comparable to SN 2011fe, the detection of
gamma-ray emission should be easily possible with an instrument like
GRIPS\@. Our simulation results show that important constraints on
three-dimensional explosion models can be drawn immediately. Since the
limited sensitivity of GRIPS below $300\,\mathrm{keV}$ hinders a
thorough examination of this energy range for sources at larger
distances, the $^{56}\mathrm{Ni}$ lines at $0.158\,\mathrm{MeV}$ and
$0.270\,\mathrm{MeV}$, which can serve as important diagnostic tools
(see also section~\ref{sec:spectra}), are not so easily accessible.
This problem can be cured by combining the gamma-ray data of GRIPS
with the data of the likewise proposed ASTRO-H mission
\citep[cf.][]{Takahashi2010} that covers the respective energy range
with sufficient sensitivity. A recent study concerning the detection
capabilities of ASTRO-H shows that distance limits comparable to those
of the GRIPS mission can be reached \citep{Maeda2012}. The realization
of two missions with sensitivities and photon energy coverages like
GRIPS and ASTRO-H in the near future would therefore offer
unprecedented detection prospects for the gamma-ray emission of SNe~Ia
and equally support the efforts in reducing the parameter space of
current three-dimensional model concepts. Especially the degeneracy 
concerning optical observables of different explosion models for the 
recent explosion event SN 2011fe \citep[cf.][]{Roepke2012} could be 
certainly removed by a study of the corresponding gamma-ray emission 
as shown in the previous paragraphs.

\section{Conclusions}
Using three-dimensional simulations of a delayed
detonation of a Chandrasekhar-mass WD and a violent merger of two WDs
as test cases, we show that the calculated gamma-ray spectra are
well-suited to draw conclusions on the different ejecta structure of
the two models, but viewing angle effects always have to be taken into
account. The direct link of gamma-ray spectra to the abundance
of $^{56}\mathrm{Ni}$ -- the radioactive isotope that powers the
radiation in all other wavelength bands -- and their reduced
complexity due to their straightforward connection to fundamental 
physical processes make them a
promising utility that can be used in a complementary way to other
measurements. Being mainly sensitive to the column density of the material
above the emission region, the gamma-ray emission of SNe Ia is well suited
to study the composition as well as the total mass of the ejecta.
Our analysis of different diagnostic tools such as line 
and hardness ratios demonstrates that especially hardness ratios offer 
the best prospects to distinguish the two models at further distances. 
We also underline the value of following the evolution of gamma-ray emission 
over an extended period of time: Even low-resolution spectra of distant 
explosion events lead to characteristic light curve shapes that allow a 
discrimination of the two models. If the flux maximum can be obtained
to a precision of about $5\,\mathrm{days}$, bolometric measurements are 
sufficient to discriminate the two models. 
The prospects of success of course depend 
on the future development in the sector of detection instruments. 
Concerning the sensitivities in the hard X-ray
as well as in the soft gamma-ray range, a major step forward will be
done with the planned ASTRO-H satellite and the proposed GRIPS
mission, allowing for observations of gamma-ray emission from SNe Ia
up to $20\,\mathrm{Mpc}$ according to conservative estimates. Although
a spectral resolution of all line features will only be provided for nearby
explosion events, our simulated GRIPS observations 
show that hardness ratio and light curve measurements can discriminate 
our models up to source distances of 10 to $16\,\mathrm{Mpc}$, dependent 
on the applied exposure time. This enhances 
the number of target candidates in the MeV energy range significantly 
and will allow for more meaningful gamma-ray emission studies of SNe Ia 
than up to now, including potentially unique opportunities for model 
discrimination.

\begin{acknowledgements}
  This work was in parts funded by the Deutsche Forschungs\-gemeinschaft (DFG)
  through the graduate school on ``Theoretical Astrophysics and Particle
Physics'' (GRK
  1147). The work of F.~K.~R{\"o}pke and M.~Fink was supported by the Deutsche
Forschungsgemeinschaft 
  via the Emmy Noether Program (RO 3676/1-1) and by the ARCHES prize of the
German Federal 
  Ministry of Education and Research (BMBF). Funding for 
  collaboration was provided by the DAAD/Go8 German-Australian
  exchange program. 
  The work of A.~Ulyanov was supported under ESA's Strategic Initiative
  AO/1-6418/10/NL/Cbi. 
  K.~Maeda recieved support from the World Premier International 
  Research Center Initiative (WPI Initiative), MEXT, Japan. 
  The simulations presented in this work were carried out at the 
  J\"{u}lich Supercomputing Centre (Germany) as part 
  of the projects HMU14 and PRA026 within the Partnership for Advanced 
  Computing in Europe (PRACE).
  The GRIPS/GRM model used in this work was developed by
  A.~Zoglauer. The GRIPS simulations were performed using the UCD
  Pascal cluster funded through the SFI Equipment Grant for High-Performance
  Computing Cluster.
  This work was supported by the NCI National Facility at the 
  Australian National University.
\end{acknowledgements}

\bibliographystyle{aa} 
\balance
\bibliography{lines}

\begin{thebibliography}{53}
\expandafter\ifx\csname natexlab\endcsname\relax\def\natexlab#1{#1}\fi

\bibitem[{{Agostinelli} {et~al.}(2003){Agostinelli}, {Allison}, {Amako},
  {Apostolakis}, {Araujo}, {Arce}, {Asai}, {Axen}, {Banerjee}, {Barrand},
  {Behner}, {Bellagamba}, {Boudreau}, {Broglia}, {Brunengo}, {Burkhardt},
  {Chauvie}, {Chuma}, {Chytracek}, {Cooperman}, {Cosmo}, {Degtyarenko},
  {dell'Acqua}, {Depaola}, {Dietrich}, {Enami}, {Feliciello}, {Ferguson},
  {Fesefeldt}, {Folger}, {Foppiano}, {Forti}, {Garelli}, {Giani},
  {Giannitrapani}, {Gibin}, {G{\'o}mez Cadenas}, {Gonz{\'a}lez}, {Gracia
  Abril}, {Greeniaus}, {Greiner}, {Grichine}, {Grossheim}, {Guatelli},
  {Gumplinger}, {Hamatsu}, {Hashimoto}, {Hasui}, {Heikkinen}, {Howard},
  {Ivanchenko}, {Johnson}, {Jones}, {Kallenbach}, {Kanaya}, {Kawabata},
  {Kawabata}, {Kawaguti}, {Kelner}, {Kent}, {Kimura}, {Kodama}, {Kokoulin},
  {Kossov}, {Kurashige}, {Lamanna}, {Lamp{\'e}n}, {Lara}, {Lefebure}, {Lei},
  {Liendl}, {Lockman}, {Longo}, {Magni}, {Maire}, {Medernach}, {Minamimoto},
  {Mora de Freitas}, {Morita}, {Murakami}, {Nagamatu}, {Nartallo}, {Nieminen},
  {Nishimura}, {Ohtsubo}, {Okamura}, {O'Neale}, {Oohata}, {Paech}, {Perl},
  {Pfeiffer}, {Pia}, {Ranjard}, {Rybin}, {Sadilov}, {di Salvo}, {Santin},
  {Sasaki}, {Savvas}, {Sawada}, {Scherer}, {Sei}, {Sirotenko}, {Smith},
  {Starkov}, {Stoecker}, {Sulkimo}, {Takahata}, {Tanaka}, {Tcherniaev}, {Safai
  Tehrani}, {Tropeano}, {Truscott}, {Uno}, {Urban}, {Urban}, {Verderi},
  {Walkden}, {Wander}, {Weber}, {Wellisch}, {Wenaus}, {Williams}, {Wright},
  {Yamada}, {Yoshida}, \& {Zschiesche}}]{Agostinelli2003}
{Agostinelli}, S., {Allison}, J., {Amako}, K., {et~al.} 2003, Nuclear
  Instruments and Methods in Physics Research A, 506, 250

\bibitem[{{Ajello} {et~al.}(2008){Ajello}, {Greiner}, {Sato}, {Willis},
  {Kanbach}, {Strong}, {Diehl}, {Hasinger}, {Gehrels}, {Markwardt}, \&
  {Tueller}}]{Ajello2008}
{Ajello}, M., {Greiner}, J., {Sato}, G., {et~al.} 2008, \apj, 689, 666

\bibitem[{{Ambwani} \& {Sutherland}(1988)}]{Ambwani1988}
{Ambwani}, K. \& {Sutherland}, P. 1988, \apj, 325, 820

\bibitem[{{Blondin} {et~al.}(2011){Blondin}, {Mandel}, \&
  {Kirshner}}]{Blondin2011}
{Blondin}, S., {Mandel}, K.~S., \& {Kirshner}, R.~P. 2011, \aap, 526, A81+

\bibitem[{{Boggs}(2006)}]{Boggs2006}
{Boggs}, S.~E. 2006, \nar, 50, 604

\bibitem[{{Burrows} \& {The}(1990)}]{Burrows1990}
{Burrows}, A. \& {The}, L.-S. 1990, \apj, 360, 626

\bibitem[{{Chan} \& {Lingenfelter}(1990)}]{Chan1990}
{Chan}, K.-W. \& {Lingenfelter}, E.~R. 1990, in International Cosmic Ray
  Conference, Vol.~1, International Cosmic Ray Conference, 101

\bibitem[{{Chan} \& {Lingenfelter}(1988)}]{Chan1988}
{Chan}, K.~W. \& {Lingenfelter}, R.~E. 1988, in American Institute of Physics
  Conference Series, Vol. 170, Nuclear Spectroscopy of Astrophysical Sources,
  ed. N.~{Gehrels} \& G.~H. {Share}, 110--115

\bibitem[{{Chan} \& {Lingenfelter}(1991)}]{Chan1991}
{Chan}, K.~W. \& {Lingenfelter}, R.~E. 1991, \apj, 368, 515

\bibitem[{{Clayton}(1974)}]{Clayton1974}
{Clayton}, D.~D. 1974, \apj, 188, 155

\bibitem[{{Clayton} {et~al.}(1969){Clayton}, {Colgate}, \&
  {Fishman}}]{Clayton1969}
{Clayton}, D.~D., {Colgate}, S.~A., \& {Fishman}, G.~J. 1969, \apj, 155, 75

\bibitem[{{Dan} {et~al.}(2011){Dan}, {Rosswog}, {Guillochon}, \&
  {Ramirez-Ruiz}}]{Dan2011}
{Dan}, M., {Rosswog}, S., {Guillochon}, J., \& {Ramirez-Ruiz}, E. 2011, \apj,
  737, 89

\bibitem[{{G{\'o}mez-Gomar} {et~al.}(1998){G{\'o}mez-Gomar}, {Isern}, \&
  {Jean}}]{Gomez-Gomar1998}
{G{\'o}mez-Gomar}, J., {Isern}, J., \& {Jean}, P. 1998, \mnras, 295, 1

\bibitem[{{Greiner} {et~al.}(2009){Greiner}, {Iyudin}, {Kanbach}, {Zoglauer},
  {Diehl}, {Ryde}, {Hartmann}, {Kienlin}, {McBreen}, {Ajello}, {Bagoly},
  {Balasz}, {Barbiellini}, {Bellazini}, {Bezrukov}, {Bisikalo},
  {Bisnovaty-Kogan}, {Boggs}, {Bykov}, {Cherepashuk}, {Chernenko}, {Collmar},
  {DiCocco}, {Dr{\"o}ge}, {Gierlik}, {Hanlon}, {Horvath}, {Hudec}, {Kiener},
  {Labanti}, {Langer}, {Larsson}, {Lichti}, {Lipunov}, {Lubsandorgiev},
  {Majczyna}, {Mannheim}, {Marcinkowski}, {Marisaldi}, {McBreen}, {Meszaros},
  {Orlando}, {Panasyuk}, {Pearce}, {Pian}, {Poleschuk}, {Pollo}, {Pozanenko},
  {Savaglio}, {Shustov}, {Strong}, {Svertilov}, {Tatischeff}, {Uvarov},
  {Varshalovich}, {Wunderer}, {Wrochna}, {Zabrodskij}, \&
  {Zeleny}}]{Greiner2009}
{Greiner}, J., {Iyudin}, A., {Kanbach}, G., {et~al.} 2009, Experimental
  Astronomy, 23, 91

\bibitem[{{Greiner} {et~al.}(2012){Greiner}, {Mannheim}, {Aharonian}, {Ajello},
  {Balasz}, {Barbiellini}, {Bellazzini}, {Bishop}, {Bisnovatij-Kogan}, {Boggs},
  {Bykov}, {DiCocco}, {Diehl}, {Els{\"a}sser}, {Foley}, {Fransson}, {Gehrels},
  {Hanlon}, {Hartmann}, {Hermsen}, {Hillebrandt}, {Hudec}, {Iyudin}, {Jose},
  {Kadler}, {Kanbach}, {Klamra}, {Kiener}, {Klose}, {Kreykenbohm}, {Kuiper},
  {Kylafis}, {Labanti}, {Langanke}, {Langer}, {Larsson}, {Leibundgut}, {Laux},
  {Longo}, {Maeda}, {Marcinkowski}, {Marisaldi}, {McBreen}, {McBreen},
  {Meszaros}, {Nomoto}, {Pearce}, {Peer}, {Pian}, {Prantzos}, {Raffelt},
  {Reimer}, {Rhode}, {Ryde}, {Schmidt}, {Silk}, {Shustov}, {Strong}, {Tanvir},
  {Thielemann}, {Tibolla}, {Tierney}, {Tr{\"u}mper}, {Varshalovich}, {Wilms},
  {Wrochna}, {Zdziarski}, \& {Zoglauer}}]{Greiner2012}
{Greiner}, J., {Mannheim}, K., {Aharonian}, F., {et~al.} 2012, Experimental
  Astronomy, 34, 551

\bibitem[{{Gruber} {et~al.}(1999){Gruber}, {Matteson}, {Peterson}, \&
  {Jung}}]{Gruber1999}
{Gruber}, D.~E., {Matteson}, J.~L., {Peterson}, L.~E., \& {Jung}, G.~V. 1999,
  \apj, 520, 124

\bibitem[{{Hillebrandt} \& {Niemeyer}(2000)}]{Hillebrandt2000}
{Hillebrandt}, W. \& {Niemeyer}, J.~C. 2000, \araa, 38, 191

\bibitem[{{H{\"o}flich}(2002)}]{Hoeflich2002}
{H{\"o}flich}, P. 2002, \nar, 46, 475

\bibitem[{{H{\"o}flich} {et~al.}(1992){H{\"o}flich}, {Khokhlov}, \&
  {M{\"u}ller}}]{Hoeflich1992}
{H{\"o}flich}, P., {Khokhlov}, A., \& {M{\"u}ller}, E. 1992, \aap, 259, 549

\bibitem[{{H{\"o}flich} {et~al.}(1998){H{\"o}flich}, {Wheeler}, \&
  {Khokhlov}}]{Hoeflich1998}
{H{\"o}flich}, P., {Wheeler}, J.~C., \& {Khokhlov}, A. 1998, \apj, 492, 228

\bibitem[{{Isern} {et~al.}(2008){Isern}, {Bravo}, \& {Hirschmann}}]{Isern2008}
{Isern}, J., {Bravo}, E., \& {Hirschmann}, A. 2008, \nar, 52, 377

\bibitem[{{Kasen} {et~al.}(2009){Kasen}, {R{\"o}pke}, \& {Woosley}}]{Kasen2009}
{Kasen}, D., {R{\"o}pke}, F.~K., \& {Woosley}, S.~E. 2009, \nat, 460, 869

\bibitem[{{Khokhlov}(1991)}]{Khokhlov1991}
{Khokhlov}, A.~M. 1991, \aap, 245, 114

\bibitem[{{Kromer} \& {Sim}(2009)}]{Kromer2009}
{Kromer}, M. \& {Sim}, S.~A. 2009, \mnras, 398, 1809

\bibitem[{{Kromer} {et~al.}(2010){Kromer}, {Sim}, {Fink}, {R{\"o}pke},
  {Seitenzahl}, \& {Hillebrandt}}]{Kromer2010}
{Kromer}, M., {Sim}, S.~A., {Fink}, M., {et~al.} 2010, \apj, 719, 1067

\bibitem[{{Kumagai} \& {Nomoto}(1997)}]{Kumagai1997}
{Kumagai}, S. \& {Nomoto}, K. 1997, in NATO ASIC Proc. 486: Thermonuclear
  Supernovae, ed. P.~{Ruiz-Lapuente}, R.~{Canal}, \& J.~{Isern}, 515

\bibitem[{{Lucy}(1999)}]{Lucy1999}
{Lucy}, L.~B. 1999, \aap, 344, 282

\bibitem[{{Lucy}(2005)}]{Lucy2005}
{Lucy}, L.~B. 2005, \aap, 429, 19

\bibitem[{{Maeda} {et~al.}(2012){Maeda}, {Terada}, {Kasen}, {R{\"o}pke},
  {Bamba}, {Diehl}, {Nomoto}, {Kromer}, {Seitenzahl}, {Yamaguchi}, {Tamagawa},
  \& {Hillebrandt}}]{Maeda2012}
{Maeda}, K., {Terada}, Y., {Kasen}, D., {et~al.} 2012, submitted

\bibitem[{{Mazzali} {et~al.}(2007){Mazzali}, {R{\"o}pke}, {Benetti}, \&
  {Hillebrandt}}]{Mazzali2007}
{Mazzali}, P.~A., {R{\"o}pke}, F.~K., {Benetti}, S., \& {Hillebrandt}, W. 2007,
  Science, 315, 825

\bibitem[{{Milne} {et~al.}(2004){Milne}, {Hungerford}, {Fryer}, {Evans},
  {Urbatsch}, {Boggs}, {Isern}, {Bravo}, {Hirschmann}, {Kumagai}, {Pinto}, \&
  {The}}]{Milne2004}
{Milne}, P.~A., {Hungerford}, A.~L., {Fryer}, C.~L., {et~al.} 2004, \apj, 613,
  1101

\bibitem[{{Mizuno} {et~al.}(2004){Mizuno}, {Kamae}, {Godfrey}, {Handa},
  {Thompson}, {Lauben}, {Fukazawa}, \& {Ozaki}}]{Mizuno2004}
{Mizuno}, T., {Kamae}, T., {Godfrey}, G., {et~al.} 2004, \apj, 614, 1113

\bibitem[{{M{\"u}ller} {et~al.}(1991){M{\"u}ller}, {H{\"o}flich}, \&
  {Khokhlov}}]{Mueller1991}
{M{\"u}ller}, E., {H{\"o}flich}, P., \& {Khokhlov}, A. 1991, \aap, 249, L1

\bibitem[{{Pakmor} {et~al.}(2012{\natexlab{a}}){Pakmor}, {Edelmann},
  {R{\"o}pke}, \& {Hillebrandt}}]{Pakmor2012b}
{Pakmor}, R., {Edelmann}, P., {R{\"o}pke}, F.~K., \& {Hillebrandt}, W.
  2012{\natexlab{a}}, \mnras, 424, 2222

\bibitem[{{Pakmor} {et~al.}(2011){Pakmor}, {Hachinger}, {R{\"o}pke}, \&
  {Hillebrandt}}]{Pakmor2011}
{Pakmor}, R., {Hachinger}, S., {R{\"o}pke}, F.~K., \& {Hillebrandt}, W. 2011,
  \aap, 528, A117

\bibitem[{{Pakmor} {et~al.}(2010){Pakmor}, {Kromer}, {R{\"o}pke}, {Sim},
  {Ruiter}, \& {Hillebrandt}}]{Pakmor2010}
{Pakmor}, R., {Kromer}, M., {R{\"o}pke}, F.~K., {et~al.} 2010, \nat, 463, 61

\bibitem[{{Pakmor} {et~al.}(2012{\natexlab{b}}){Pakmor}, {Kromer},
  {Taubenberger}, {Sim}, {R{\"o}pke}, \& {Hillebrandt}}]{Pakmor2012a}
{Pakmor}, R., {Kromer}, M., {Taubenberger}, S., {et~al.} 2012{\natexlab{b}},
  \apjl, 747, L10

\bibitem[{{Reinecke} {et~al.}(1999){Reinecke}, {Hillebrandt}, {Niemeyer},
  {Klein}, \& {Gr{\"o}bl}}]{Reinecke1999}
{Reinecke}, M., {Hillebrandt}, W., {Niemeyer}, J.~C., {Klein}, R., \&
  {Gr{\"o}bl}, A. 1999, \aap, 347, 724

\bibitem[{{R{\"o}pke} {et~al.}(2006){R{\"o}pke}, {Gieseler}, {Reinecke},
  {Travaglio}, \& {Hillebrandt}}]{Roepke2006b}
{R{\"o}pke}, F.~K., {Gieseler}, M., {Reinecke}, M., {Travaglio}, C., \&
  {Hillebrandt}, W. 2006, \aap, 453, 203

\bibitem[{{R{\"o}pke} \& {Hillebrandt}(2005)}]{Roepke2005a}
{R{\"o}pke}, F.~K. \& {Hillebrandt}, W. 2005, \aap, 431, 635

\bibitem[{{R{\"o}pke} {et~al.}(2012){R{\"o}pke}, {Kromer}, {Seitenzahl},
  {Pakmor}, {Sim}, {Taubenberger}, {Ciaraldi-Schoolmann}, {Hillebrandt},
  {Aldering}, {Antilogus}, {Baltay}, {Benitez-Herrera}, {Bongard}, {Buton},
  {Canto}, {Cellier-Holzem}, {Childress}, {Chotard}, {Copin}, {Fakhouri},
  {Fink}, {Fouchez}, {Gangler}, {Guy}, {Hachinger}, {Hsiao}, {Chen},
  {Kerschhaggl}, {Kowalski}, {Nugent}, {Paech}, {Pain}, {Pecontal}, {Pereira},
  {Perlmutter}, {Rabinowitz}, {Rigault}, {Runge}, {Saunders}, {Smadja},
  {Suzuki}, {Tao}, {Thomas}, {Tilquin}, \& {Wu}}]{Roepke2012}
{R{\"o}pke}, F.~K., {Kromer}, M., {Seitenzahl}, I.~R., {et~al.} 2012, \apjl,
  750, L19

\bibitem[{{R{\"o}pke} \& {Niemeyer}(2007)}]{Roepke2007}
{R{\"o}pke}, F.~K. \& {Niemeyer}, J.~C. 2007, \aap, 464, 683

\bibitem[{{Schmidt} {et~al.}(2006){Schmidt}, {Niemeyer}, {Hillebrandt}, \&
  {R{\"o}pke}}]{Schmidt2006}
{Schmidt}, W., {Niemeyer}, J.~C., {Hillebrandt}, W., \& {R{\"o}pke}, F.~K.
  2006, \aap, 450, 283

\bibitem[{{Seitenzahl} {et~al.}(2013){Seitenzahl}, {Ciaraldi-Schoolmann},
  {R{\"o}pke}, {Fink}, {Hillebrandt}, {Kromer}, {Pakmor}, {Ruiter}, {Sim}, \&
  {Taubenberger}}]{Seitenzahl2013}
{Seitenzahl}, I.~R., {Ciaraldi-Schoolmann}, F., {R{\"o}pke}, F.~K., {et~al.}
  2013, \mnras, 429, 1156

\bibitem[{{Seitenzahl} {et~al.}(2009){Seitenzahl}, {Meakin}, {Townsley},
  {Lamb}, \& {Truran}}]{Seitenzahl2009}
{Seitenzahl}, I.~R., {Meakin}, C.~A., {Townsley}, D.~M., {Lamb}, D.~Q., \&
  {Truran}, J.~W. 2009, \apj, 696, 515

\bibitem[{{Seitenzahl} {et~al.}(2010){Seitenzahl}, {R{\"o}pke}, {Fink}, \&
  {Pakmor}}]{Seitenzahl2010}
{Seitenzahl}, I.~R., {R{\"o}pke}, F.~K., {Fink}, M., \& {Pakmor}, R. 2010,
  \mnras, 407, 2297

\bibitem[{{Sim}(2007)}]{Sim2007}
{Sim}, S.~A. 2007, \mnras, 375, 154

\bibitem[{{Sim} \& {Mazzali}(2008)}]{Sim2008}
{Sim}, S.~A. \& {Mazzali}, P.~A. 2008, \mnras, 385, 1681

\bibitem[{{Springel}(2005)}]{Springel2005}
{Springel}, V. 2005, \mnras, 364, 1105

\bibitem[{{Summa} {et~al.}(2011){Summa}, {Els{\"a}sser}, \&
  {Mannheim}}]{Summa2011}
{Summa}, A., {Els{\"a}sser}, D., \& {Mannheim}, K. 2011, \aap, 533, A13

\bibitem[{{Takahashi} {et~al.}(2010){Takahashi}, {Mitsuda}, {Kelley},
  {Aharonian}, {Akimoto}, {Allen}, {Anabuki}, {Angelini}, {Arnaud}, {Awaki},
  {Bamba}, {Bando}, {Bautz}, {Blandford}, {Boyce}, {Brown}, {Chernyakova},
  {Coppi}, {Costantini}, {Cottam}, {Crow}, {de Plaa}, {de Vries}, {den Herder},
  {Dipirro}, {Done}, {Dotani}, {Ebisawa}, {Enoto}, {Ezoe}, {Fabian},
  {Fujimoto}, {Fukazawa}, {Funk}, {Furuzawa}, {Galeazzi}, {Gandhi}, {Gendreau},
  {Gilmore}, {Haba}, {Hamaguchi}, {Hatsukade}, {Hayashida}, {Hiraga}, {Hirose},
  {Hornschemeier}, {Hughes}, {Hwang}, {Iizuka}, {Ishibashi}, {Ishida},
  {Ishimura}, {Ishisaki}, {Isobe}, {Ito}, {Iwata}, {Kaastra}, {Kallman},
  {Kamae}, {Katagiri}, {Kataoka}, {Katsuda}, {Kawaharada}, {Kawai}, {Kawasaki},
  {Khangaluyan}, {Kilbourne}, {Kinugasa}, {Kitamoto}, {Kitayama}, {Kohmura},
  {Kokubun}, {Kosaka}, {Kotani}, {Koyama}, {Kubota}, {Kunieda}, {Laurent},
  {Lebrun}, {Limousin}, {Loewenstein}, {Long}, {Madejski}, {Maeda},
  {Makishima}, {Markevitch}, {Matsumoto}, {Matsushita}, {McCammon}, {Miller},
  {Mineshige}, {Minesugi}, {Miyazawa}, {Mizuno}, {Mori}, {Mori}, {Mukai},
  {Murakami}, {Murakami}, {Mushotzky}, {Nakagawa}, {Nakagawa}, {Nakajima},
  {Nakamori}, {Nakazawa}, {Namba}, {Nomachi}, {O'Dell}, {Ogawa}, {Ogawa},
  {Ogi}, {Ohashi}, {Ohno}, {Ohta}, {Okajima}, {Ota}, {Ozaki}, {Paerels},
  {Paltani}, {Parmar}, {Petre}, {Pohl}, {Porter}, {Ramsey}, {Reynolds},
  {Sakai}, {Sambruna}, {Sato}, {Sato}, {Serlemitsos}, {Shida}, {Shimada},
  {Shinozaki}, {Shirron}, {Smith}, {Sneiderman}, {Soong}, {Stawarz}, {Sugita},
  {Szymkowiak}, {Tajima}, {Takahashi}, {Takei}, {Tamagawa}, {Tamura}, {Tamura},
  {Tanaka}, {Tanaka}, {Tanaka}, {Tashiro}, {Tawara}, {Terada}, {Terashima},
  {Tombesi}, {Tomida}, {Tozuka}, {Tsuboi}, {Tsujimoto}, {Tsunemi}, {Tsuru},
  {Uchida}, {Uchiyama}, {Uchiyama}, {Ueda}, {Uno}, {Urry}, {Watanabe}, {White},
  {Yamada}, {Yamaguchi}, {Yamaoka}, {Yamasaki}, {Yamauchi}, {Yamauchi},
  {Yatsu}, {Yonetoku}, \& {Yoshida}}]{Takahashi2010}
{Takahashi}, T., {Mitsuda}, K., {Kelley}, R., {et~al.} 2010, in Society of
  Photo-Optical Instrumentation Engineers (SPIE) Conference Series, Vol. 7732,
  Society of Photo-Optical Instrumentation Engineers (SPIE) Conference Series

\bibitem[{{Travaglio} {et~al.}(2004){Travaglio}, {Hillebrandt}, {Reinecke}, \&
  {Thielemann}}]{Travaglio2004}
{Travaglio}, C., {Hillebrandt}, W., {Reinecke}, M., \& {Thielemann}, F.-K.
  2004, \aap, 425, 1029

\bibitem[{{Zoglauer} {et~al.}(2006){Zoglauer}, {Andritschke}, \&
  {Schopper}}]{Zoglauer2006}
{Zoglauer}, A., {Andritschke}, R., \& {Schopper}, F. 2006, \nar, 50, 629

\end{thebibliography}

\end{document}